\documentclass[12pt]{article}

\usepackage{amsmath, amsthm}
\usepackage{multirow}
\usepackage{times}
\usepackage{natbib}
\usepackage[margin=1.0in]{geometry}
\newtheorem{assumption}{Assumption}[section]
\newtheorem{definition}{Definition}[section]

\AtEndDocument{\bigskip{\footnotesize%
  \textsc{Panos Toulis}, Department of Statistics, Harvard University \par  
  \textit{E-mail address:} \texttt{ptoulis@fas.harvard.edu} \par
  \addvspace{\medskipamount}
  \textsc{David C. Parkes}, SEAS, Harvard University\par  
  \textit{E-mail address:} \texttt{parkes@eecs.harvard.edu} \par
}}

\usepackage{setspace}
\usepackage[plain,noend]{algorithm2e}
\usepackage{graphicx}
\usepackage{caption}
\usepackage{subcaption}
\usepackage{authblk}

\begin{document}
\title{Long-term causal effects of economic mechanisms on agent incentives}
\author{Panos Toulis and David C. Parkes}
\maketitle
\begin{abstract}
  Economic \emph{mechanisms} administer the allocation of resources to
  interested agents based on their self-reported \emph{types}.  One
  objective in \emph{mechanism design} is to design a
  \emph{strategyproof} process so that no agent will have an
  incentive to misreport its type.  However, typical analyses of the
  incentives properties of mechanisms operate under strong, usually
  untestable assumptions. Empirical, data-oriented approaches are, at
  best, under-developed.
Furthermore, mechanism/policy evaluation methods usually ignore the dynamic
nature of a multi-agent system and are thus inappropriate for estimating long-term 
effects.
We introduce the problem of estimating the \emph{causal effects} of mechanisms on  incentives
and frame it under the Rubin causal framework \citep{rubin74, rubin78}.
This raises unique technical challenges since the outcome of interest (agent truthfulness) is
confounded with strategic interactions and, interestingly, is typically never observed under any mechanism. 
We develop a methodology to estimate such causal effects that using a prior
that is based on a strategic equilibrium model. 
Working on the domain of \emph{kidney exchanges},
we show how to apply our methodology to estimate causal effects of kidney
allocation mechanisms on hospitals' incentives. Our results demonstrate that the use 
of game-theoretic prior captures the dynamic nature of the kidney exchange multi-agent system
and shrinks the estimates towards long-term effects, thus improving upon 
typical methods that completely ignore agents' strategic behavior.

\vfill
\end{abstract}

\section{Introduction}
A mechanism defines the rules that are used to determine the
allocation of resources to interested parties in an economic
transaction.  For example, an online ad auction determines the winners
of the advertisement slots and the appropriate payments, based on
advertisers' reports (bids).  In designing mechanisms, a key
requirement is to have good incentives properties, so that agents have
no incentive to misreport their valuations.  Such strategyproof
mechanisms are appealing for being strategically simple for agents,
and can lead to desirable outcomes because this agent behavior can be
anticipated and leveraged for good effect.
%
% and more efficient as, naturally, rational
%agents undereport their valuations when lying \citep{lubin2012}.

There are a few general procedures for devising strategyproof
mechanisms.  One key idea is to determine the payment of an agent $i$
as a function of the reports of all agents excluding $i$. Along with
allocating the desired resources to this agent given the prices it
faces, the intuition is that agent $i$ will have no incentive to
misreport since this will not affect the utility given fixed reports
from others. This idea underlies the Vickrey auction \citep{vickrey61}
and its generalization in the Vickrey-Clarke-Groves (VCG) mechanism
\citep{clarke71, groves73}. For example, in the Vickrey auction, the highest bidder wins
the item but pays the second highest bid and thus, there is no incentive for the highest bidder
to reduce the initial bid.

However, and even when strategyproof mechanisms can in principle be
designed, theoretical analyses of their incentive properties rely
critically on assumptions that are strong or untestable in practice.
Typical assumptions include: 
 no \emph{collusion} among agents;
(ii)
the {\em rationality} of participants;
(iii) that the types (such as
valuations) of agents are correctly modeled (e.g., they are private
values and don't depend on information of others); and 
(iv) that the
strategic interactions have been correctly modeled. 
But without getting these assumptions correct, the incentive
properties of a mechanism will not be as desired.  For example, if
participants in a single-item second price (Vickrey) auction can
collude then one bidder can submit a high bid while others withhold
their bids. In another example, if the problem is truly multi-round
then participants have a new incentive to shave down their bids in
order to get the best price for an item across time.

In this light, there remains a large opportunity for empirical methods
in the design of mechanisms, especially in estimating the \emph{causal
  effects} of mechanisms on incentives, and other outcomes of interest
(such as welfare, revenue and so forth.) Across many online platforms
such as ad auction platforms, one wants to be able to make changes to
the design across subsets of the population and be able to estimate
the effect of these design decisions on economic properties if one was
to run a single design on the whole population vs run some other
design.  In this paper we adopt the Rubin causal framework
\citep{rubin74, rubin78} using the potential outcomes notation.  
Our goal is to estimate the causal effects of mechanisms on the incentives of agents, 
after agents have been randomly assigned to participate in one mechanism (viewed as a treatment)
and their reports have been observed. This raises unique technical challenges
since the outcome of interest (agent truthfulness) is typically never observed
under any mechanism. Furthermore,  we assume that 
data collection (agent reports) happens before the system has reached an equilbrium 
and so, we are interested in a methodology that will strike a sensible balance between
observed data and equilbrium considerations.

There is a developing body of work on experimention of online,
socio-economic systems, but it is not developed within the potential
outcomes framework and has not studied the special question of causal
analysis in regard to incentive properties. In related work, 
large field experiment conducted at Yahoo! in 2008, aimed to estimate the
effects of increased reserve prices on keyword revenue \citep{ostrovsky2011}.
The applied method was to use a ``diff-in-diffs" estimator that completely ignores all 
aforementioned subtleties. Other work aims to estimate the effects of interventions
in a machine learning model underlying a mechanism (e.g. \citep{bottou2012}), but 
the methods are usually predictive (i.e. predict all missing outcomes through a model
based on one intervention and the observed outcomes). Equilibrium effects for causal inference
has first been proposed in the econometric literature \citep{heckman05}.
However, no general methodology has been proposed for the estimation of long-term causal effects 
of policies/mechanisms.

Other work has considered the empirical design of mechanisms in
settings where the goal of strategyproof design is unachievable in
combination with other desirable properties, or cannot be supported
from an analytical framework~\citep{lubin09}. This appeals to the
divergence between distributions over payments (or payoffs) in an
incentive-aligned mechanism and distributions over payments (or
payoffs) in another candidate design, with the view to finding the
optimal mechanism through online search. But this work does not adopt
a causal approach, but rather assumes the ability to switch the entire
population through alternate designs and thus does not have the
difficulty of estimating counterfactuals. Moreover, the work does not
adopt our viewpoint of looking to make inferences from empirical
frequences about reported types about the incentive properties of the
mechanism.

% mathcal for agents, types etc.
\newcommand{\m}[2]{\mathcal{#1}_{#2}}
\newcommand{\A}[1]{\m{A}{#1}}
\renewcommand{\S}{\m{S}{}}
\newcommand{\M}[1]{\mathcal{M}_{#1}}
\newcommand{\Mt}{\M{1}}
\newcommand{\Mc}{\M{0}}

% Mechanisms
\renewcommand{\M}[1]{\m{M}{#1}}

% strategies={truthful, deviation}
\newcommand{\st}{t}
\newcommand{\sd}{d}

% potential outcomes
\newcommand{\yi}[1]{y_i(#1)}
\newcommand{\Yi}[2]{Y_i(#1, #2)}
\newcommand{\Yizt}{\Yi{z}{t}}
\newcommand{\Zi}{Z_i}
\newcommand{\zi}{\Zi}
\newcommand{\Ri}[2]{R_i(#1, #2)}
\newcommand{\Rizt}{\Ri{z}{t}}
\newcommand{\ri}[1]{r_i(#1)}  %% report of i
\newcommand{\typei}{\theta_{it}}
\newcommand{\typespace}{\Theta}
\newcommand{\binary}{\{0, 1\}}

%% Vectors
\newcommand{\Z}{\mathbf{Z}}
\newcommand{\Y}[2]{\mathbf{Y}_{#1}(#2)}
\newcommand{\R}[2]{\mathbf{R}_{#1}({#2})}
\newcommand{\Yt}{\Y{1}{t}}
\newcommand{\Ytreal}{\mathbf{Y}_{1}^{\text{real}}(t)}
\newcommand{\Ytrealo}{\mathbf{Y}_{1}^{\text{real}}(t)}
\newcommand{\Ytnreal}{\mathbf{Y}_{1}^{\text{nreal}}(t)}
\newcommand{\Ytnrealo}{\mathbf{Y}_{1}^{\text{nreal}}(t)}
\newcommand{\Ytothers}[2]{\mathbf{Y}_{1,-#1}(#2)}
\newcommand{\Ycothers}[2]{\mathbf{Y}_{0,-#1}(#2)}

\newcommand{\Yc}{\Y{0}{t}}
\newcommand{\Ycreal}{\mathbf{Y}_{0}^{\text{real}}(t)}
\newcommand{\Ycrealo}{\mathbf{Y}_{0}^{\text{real}}(t)}
\newcommand{\Ycnreal}{\mathbf{Y}_{0}^{\text{nreal}}(t)}
\newcommand{\Ycnrealo}{\mathbf{Y}_{0}^{\text{nreal}}(t)}
\newcommand{\Rt}{\R{1}{t}}
\newcommand{\Rtobs}{\mathbf{R}_{1}^{obs}(t)}
\newcommand{\Rcobs}{\mathbf{R}_{0}^{obs}(t)}

\newcommand{\Rc}{\R{0}{t}}
\newcommand{\RI}{\Rtobs}
\newcommand{\RO}{\Rcobs}
\newcommand{\Yobs}[1]{\Y{#1}{obs}}

% Model
\newcommand{\ysi}{p \left (\yi{m} | \ri \right )}
\newcommand{\syi}{p \left (\ri | \yi{m}\right )}
\newcommand{\piyi}{\pi \left ( \yi{m} \right )}
\newcommand{\prior}[1]{\pi(\Y{#1}{})}
\newcommand{\hyperprior}[2]{\pi(\Y{#1}{} | #2)}

%% Brevity
\newcommand{\YI}{\Yt}
\newcommand{\YO}{\Yc}
\newcommand{\LI}{\mathcal{L}(\RI | \YI)}
\newcommand{\LO}{\mathcal{L}(\RO | \YO)}
\renewcommand{\L}{\mathcal{L}(\RI, \RO | \YO, \YI)}
\newcommand{\Lj}[1]{\mathcal{L}(\R{#1}{obs} | \Y{#1}{t})}

\def\independenT#1#2{\mathrel{\rlap{$#1#2$}\mkern2mu{#1#2}}}
\newcommand\independent{\protect\mathpalette{\protect\independenT}{\perp}}
\newcommand{\U}{\mathbf{U}}
\newcommand{\Uj}[1]{\U(\Y{j}{} ; \M{j})}
\newcommand{\bb}{\boldsymbol{\beta}}

%%%  Experiments --- Kidney Exchanges.
\newcommand{\rcm}{\M{0}}
\newcommand{\xcm}{\M{1}}
\newcommand{\E}[1]{\texttt{UCB}(#1)}
\newcommand{\GT}{\texttt{GT}}
\newcommand{\ui}{\bar{u_i}}
\newcommand{\tcb}[1]{\tau_{UCB}(#1)}

\section{Causal effects on incentives}
We consider a population of $N$ agents, indexed by $i$ in some natural ordering, and two mechanisms $\Mt$ and $\Mc$.
Each agent will be randomized to participate in one mechanism only (thus the mechanisms 
can be viewed as treatments that agents receive).
Specifically, if agent $i$ participates in $\Mt$ then $Z_i=1$ and if agent $i$ participates in $\Mc$ then $\Zi=0$.
We also consider a setting where the mechanisms are multi-round, each round being indexed by $t$, 
for $t=1, 2, \cdots T$ and a maximum number of rounds $T$.
At each round $t$ of any mechanism, each agent samples a type $\typei$ from a type space $\typespace$
according to some distribution $F$ and selects a strategy $\Yizt \in \binary$. 
Given the sampled type, agent $i$, then reports $\Rizt$ to mechanism $\mathcal{M}_z$ according to the following rule:
\begin{eqnarray}
\label{eq:po}
  \Rizt=\begin{cases}
    \typei, & \text{if $\Yizt=1$} \nonumber \\
    d(\typei) \in \typespace, & \text{ otherwise }
  \end{cases}
\end{eqnarray}

In other words, if $\Yizt=1$, agent $i$ is reporting truthfully to the mechanism
and it is deviating according to a known deviation if $\Yizt=0$.
We assume that, the distribution of true types $F$ and the function
of deviation $d(\cdot)$, are known or can be estimated from other sources\footnote{
For example, in the ad auction literature bidder valuations can be routinely estimated from the data
using empirical distribution of bids and prices \citep{athey2010, ostrovsky2011}}.
Thus, assume that $d(\typei)$ has a known distribution $G$.
The agent strategy $\Yizt$ and report $\Rizt$ at each round $t$ are the potential outcomes of interest.

We consider a completely randomized experiment and denote the entire $(N\times 1)$ assignment vector with $\Z$.
The full $(N\times 1)$ vector of potential outcomes of agent stratagies for mechanisms $\Mc$ and $\Mt$ at round $t$,
are denoted by $\Yc$ and $\Yt$ respectively. In a similar fashion, let $\Rc, \Rt$ denote the 
potential reports of agents in mechanisms $\Mc, \Mt$ respectively. Note that all 
the potential outcomes for agent strategies, $\Yc, \Yt$, are never observed and thus are considered as missing.
However, we make the following distinction: for an agent $i$ with $Z_i=z \in \binary$, 
the outcome $\Yi{z}{t}$ will be \emph{realized} but will not be observed whereas
the outcome $\Yi{1-z}{t}$ will not be realized at all. The subvector of $\Yc$
with the realized outcomes for $\Mc$ is denoted by $\Ycreal$.
Similarly, the vector of realized outcomes for $\Mt$ is denoted by $\Ytreal$. In contrast, some of the potential outcomes of the reports of agents are observed under 
the mechanisms they participate in. Let $\Rcobs$ denote the subvector of $\Rc$, for those 
agents $i$ such that $Z_i=0$, and $\Rtobs$ be the subvector of $\Rt$ for agents $i$ with $Z_i=1$.

%%  SCIENCE TABLE
\renewcommand{\arraystretch}{1.3}
\begin{table}[h]
\caption{Science table of potential outcomes. Outcomes of agent strategies $\Yizt$ are all missing
and reports $R_i(z, t)$ are only observed when $Z_i=z$.} 
\label{sample-table}
\begin{center}
\begin{tabular}{cc|cc|cc}
&  &  \multicolumn{2}{c}{$\M{0}$} &  \multicolumn{2}{c}{$\M{1}$} \\
Units & $\Z$ & $\Rc$ & $\Yc$ & $\Rt$ & $\Yt$ \\
\hline
$1$ & 0   &  $R_1(0, t)$     &     ?      &        ?     &         ? \\
$2$ & 0   & $R_2(0, t)$     &     ?      &        ?     &         ? \\
\multicolumn{2}{c}{$\cdots$} & \multicolumn{2}{c}{$\cdots$}  & \multicolumn{2}{c}{$\cdots$}  \\
$N$-$1$ & 1   & ?               & ?          & $R_{N-1}(1, t)$        & ? \\
$N$ & 1   & ?               & ?          & $R_{N}(1, t)$        & ? \\
\end{tabular}
\label{table:science}
\end{center}
\end{table}

The ``Science" table of the observed and unobserved quantities in the aforementioned experiment is 
shown in Table \ref{table:science}. We can now define the causal estimand of interest:

\begin{definition}[Causal effects on incentives]
The causal effect on incentives of mechanism $\M{1}$ over mechanism $\M{0}$ in round $t$, is
defined by:
\begin{align}
\label{def:estimand}
	\Delta(t) = \frac{1}{N} \sum_{i=1}^N \Yi{1}{t} - \Yi{0}{t}
\end{align}
The estimand $\Delta(T)$ is defined as the long-term effect on agent incentives.

\end{definition}

%% Discuss about the causal estimand.
\subsection{Discussion}
Recall that $\Yizt$ denotes the strategy of agent $i$ (1=truthful, 0=deviating) it mechanism $\M{z}$ at round $t$.
Therefore, the estimand $\Delta(t)$ defined in \eqref{def:estimand} compares the proportions of truthful agents in 
$\Mt$ compared to $\Mc$ and by definition, it holds that $\Delta(t) \in [-1, 1]$.
Other options for the definition of the estimand are available as well (e.g. median of difference)
and in general it would involve a ``contrast function" $h(\Yt, \Yc)$ that will summarize 
the difference between the two vectors. We will use this notion of contrast function throughout the rest
of this paper, but for all numerical purposes we will assume this is the difference in means as in Definition \ref{def:estimand}.

Note also that the estimand is time-dependent as we expect agents to be self-interested and adapt their strategy over repeated rounds.
The long-term effect $\Delta(T)$ is trying to capture this dynamic evolution
of agent strategies over a specified time horizon $T$ that is considered enough time for the system 
to reach an economic equilbrium. This is related to the study of \emph{equilibrium effects} in the econometric
literature \citep{heckman05}.

%%
%%
%%
%% 		CAUSAL INFERENCE			%%% 
\section{Causal Inference}
Inspection of Table \ref{table:science} reveals one major technical challenge.
Since we do not observe the actual strategies of agents (i.e., their ``truthfulness" status)
but only their reports, no potential outcomes of strategies $\Yizt$ are actually observed.
However, given strategies $\Yizt$ the potential outcomes on reports $\Rizt$ have a 
well-defined distribution\footnote{Since we operate under a completely randomized experiment, the assignment mechanism $p(\Z|\cdots)$ is unconfounded and the vector $\Z$ can be omitted for brevity.}.
\begin{align}
\label{eq:joint}
p(\YI, \YO | \RI, \RO, \Z) &  \propto \L  \nonumber \\
                         \times &  \pi(\YO, \YI)                        
\end{align}

The likelihood term $\L$ is easy to obtain based on our assumptions. 
Specifically, we have assumed that report $\Rizt$ has distribution $F$ if 
agent $i$ is truthful and has distribution $G$ if it is deviating.
 Hence:
\begin{align}
\L = & \LI \times \LO \nonumber
\end{align}

Therefore, by independence, it holds for $j \in \{0,1\}$ indexing mechanism $\mathcal{M}_j$:
\begin{eqnarray}
\label{eq:likelihood}
\mathcal{L}(\textbf{R}_j^{\text{obs}}(t) | \textbf{Y}_j^{\text{real}}(t)) = \prod_{i : \zi=j} f(\Ri{j}{t})^{\Yi{j}{t}} g(\Ri{j}{t})^{1- \Yi{j}{t}}
\end{eqnarray}

Hence, causal inference depends critically on the model of $\pi(\YO, \YI)$.
The main contribution of this paper is to consider a prior on potential outcomes that 
has a game-theoretic justification through a well-defined equilibrium model. 
The main idea is that, by doing so, we will shrink estimates from data observed 
at an early round towards the long-term effects, assuming that the equilibrium model is accurate enough to describe the dynamics of the economic system.
To illustrate our method we will compare it with a straightforward imputation method
the is based on a uniform prior. More options, such as a fully-Bayesian approach are discussed later.

\subsection{Empirical method: Imputation on uniform prior of realized outcomes}
This method, dubbed the \emph{empirical method}, serves as our baseline method and works in two steps.
First, we impute the realized (but missing outcomes) $\Ytreal, \Ycreal$ assuming a uniform prior.
Second, we impute the non-realized and missing outcomes $\Ycnreal, \Ytnreal$ through the
empirical distribution of the imputed realized outcomes. This algorithmic process (shown next) is repeated 
many times and estimates of the causal effects are used for summarization.

\begin{algorithm}[!h]
\caption{Causal inference on incentives through uniform priors. Unrealized outcomes are
	sampled from the empirical distribution of the imputed realized ones. Variable 
$n$ is the \# of samples.} 
\label{algo1}
\vspace*{5pt}
\begin{tabbing}
   \enspace Initialize $\Delta= $ array of length $n$  \\
   \enspace For $j = 1,2, \cdots n$ \\
   \qquad Impute all missing potential outcomes for strategies as follows: \\
   \qquad\qquad  $\Ytreal  \sim \mathcal{L} \left ( \textbf{R}^{\text{obs}}(1, t) | \Ytrealo \right )  $\\
 	\qquad\qquad $\Ytnreal  \sim \Ytrealo $ \hspace{20px} (use empirical distribution)\\
   \qquad\qquad $\Ycreal  \sim \mathcal{L} \left ( \textbf{R}^{\text{obs}}(0, t) | \Ycrealo \right )  $\\
 	\qquad\qquad $\Ycnreal  \sim \Ycrealo $ \hspace{20px} (use empirical distribution) \\
	\qquad $\textbf{Y}_1(t) = (\Ytnreal, \Ytreal)$ \\ 
	\qquad $\textbf{Y}_0(t) = (\Ycreal, \Ycnreal)$ \\
 	\qquad Causal effect estimate $\hat{\Delta_j}(t) = h(\textbf{Y}_1(t), \textbf{Y}_0(t))$ \\
\enspace Return $\Delta$
\end{tabbing}
\end{algorithm}

The estimate of causal effects on incentives from the empirical method is given by:
\begin{eqnarray}
\label{eq:empirical}
\widehat{\Delta(t)} = \frac{1}{n} \sum_j \hat{\Delta_j}(t) 
\end{eqnarray}

One critical implicit assumption underlying the imputation of non-realized outcomes
from realized ones, is that collective behavior is somehow ``homegeneous" in a mechanism.
For example, if 2 out of 10 agents are truthful on average, then we expect 4 agents to be truthful out of 20.

\newcommand{\Robs}{\R{}{obs}}
\newcommand{\Lobs}{\mathcal{L}(\Robs | \YO, \YI)}

%% GOOD UP TO HERE
%%
%% Game-theoretic method (prior)

\subsection{Game-theoretic method: Imputation using a game-theoretic prior}
Typical causal inference methods, such as the aforementioned one, are
usually criticized that they ignore incentives in a multi-agent system \citep{heckman05}.
However, in this work we show that the Rubin causal model can be adapted to address this issue.
Before we proceed, we make the following assumption:
\begin{assumption} [Best-response behavior]
\label{assumption:br}
Given no prior information, the potential outcomes on agent strategies 
are independent for every round i.e., $\YI \independent \YO, \forall t$.
\end{assumption}

Assumption \ref{assumption:br} can be thought as a consequence of assuming
that agents are best-responding, regardless of behaviors in other mechanisms.
This assumption would be invalid in several cases, for example, when agents 
have different propensities to be truthful or lie to a mechanism. We will offer more discussion 
later in the paper.

Assuming independence, we only need to model $\pi(\Yc)$ and, similarly, $\pi(\Yt)$.
Recall that in any mechanism, say $\Mt$, the expected utility of 
agent $i$ choosing strategy $\Yizt=y$ assuming fixed behaviors from other agents $\Ytothers{i}{t}$ 
is denoted by $u_i(y, \Ytothers{i}{t})$. Therefore the expected utility benefit 
from being truthful for agent $i$ is given by 
\begin{eqnarray}
\label{eq:utility}
\Delta u_i(\Ytothers{i}{t}) = u_i(1, \Ytothers{i}{t}) - u_i(0, \Ytothers{i}{t})
\end{eqnarray}

We adopt a \emph{quantal response equilbrium} \citep{mckelvey95, goeree03} model  in order to construct our game-theoretic prior. In specific, agent $i$ facing a vector of agent strategies, will randomize over the available actions according to a softmax rule that depends on the expected utilities. 
In specific, for a mechanism $\mathcal{M}_z$ at round $t$, agent $i$ facing fixed behaviors 
from other agents $\textbf{Y}_{z, -i}(t)$, will select to be truthful according to the probability:
\begin{eqnarray}
\label{eq:prior:game}
\pi(\Yizt =1| \textbf{Y}_{z, -i}(t)) \propto \exp \left ( \beta \cdot u_i(1, \textbf{Y}_{z, -i}(t)) \right ) 
\end{eqnarray}

Hence:
\begin{eqnarray}
\label{eq:prior:game}
\pi(\Yizt =1| \textbf{Y}_{z, -i}(t))  = \frac{1 }{1 + \exp \left ( \beta \cdot  \Delta u_i(\textbf{Y}_{z, -i}(t)) \right )} \nonumber
\end{eqnarray}

Quantal response equilbrium is a well-studied model of utility-based agent behavior that has 
been shown to converge to Nash equilibria under certain mild conditions \citep{mckelvey95}.
The choice of parameter $\beta >0$ is critical\footnote{If $\beta < 0$ that would be considered irrational since 
the agent would prefer actions with smaller expected utilities than others.}. If $\beta$ is high
then the agent has a strong preference for actions with better expected utilities. In the extreme case,
if $\beta$ is very high then the agent simply prefers the best action (so adopts a best-response strategy).
We will discuss the choice of $\beta$ in the experimental section. Note also that there are $N$ functions $u_i(\cdot)$ and each needs to be evaluated at $2^{N-1}$ points
since the outcome is binary. For large populations this computation is prohibitive. To circumvent this problem we need to make the following exchangeability assumption:
\begin{assumption} [Exchangeability among agents]
\label{assumption:xchangeability}
Agents are exchangeable so that inferences are invariant to permutations of agent labels.
\end{assumption}

The main consequence of the exchangeability assumption is that the expected utility $u_i(\cdot)$ are the same for all agents. Furthermore, the expected utility of an agent depends only on the number of truthful agents he is ``competing"
 against and so  $\Delta u_i(\textbf{Y}_{z, -i}(t)) = \Delta u(\sum_{j \ne i} Y_j(z, t))$.
In other words, there is only one function $\Delta u(\cdot)$ for each agent that needs to be evaluated at $N+1$ points (having 0 to N truthful agents in total).
Inference can now be performed through Gibbs sampling. The implementation is straightforward 
since strategy outcomes are binary. In summary, if $i$ participates in mechanism $\mathcal{M}_z$, 
then potential outcome $Y_i(z, t)$ is sampled according to the following rule: If the outcome is not-realized then only
the report outcome $R_i(z, t)$ is used (this is observed) to sample the strategy. However, if the outcome is realized, then 
both the report $R_i(z, t)$ and the vector of agent behaviors $\textbf{Y}_{z, -i}(t)$ is used to sample $\Yizt$.
The full procedure is given in Algorithm \ref{algo2}.
\begin{algorithm}[!h]
\caption{Causal inference on incentives through a game-theoretic prior. 
Inference performed though Gibbs sampler. Each sample is indexed by $j$.
Variable $n$ is the \# of samples.} 
\label{algo2}
\vspace*{5pt}
\begin{tabbing}
   \enspace Initialize $\Delta= $ array of length $n$  \\
   \enspace For $j = 1,2, \cdots n$ \\
  \qquad Initialize $\textbf{Y}_0^{(j)}(t) = \textbf{Y}_0^{(j-1)}(t) $ and $\textbf{Y}_1^{(j)}(t) = \textbf{Y}_1^{(j-1)}(t) $ \\
   \qquad $\forall i \in \{1,2,\cdots, N\}$ and $z \in \binary$: \\
 \qquad\qquad $\pi_{iz} = \left (1 + \exp \left (\beta \Delta u(\sum_{j \ne i} Y_j(z, t)) \right ) \right )^{-1}$ \\
  \qquad\qquad    $p_{iz}=\begin{cases}
    					\mathcal{L}(R_i(z, t) | Y_i^{(j)}(z, t)=1) , & \text{if $Z_i=1-z$} \nonumber \\
    					\mathcal{L}(R_i(z, t) | Y_i(z, t)=1) \times  \pi_{iz}, &  \text{if $Z_i=z$} 
  \end{cases}$ \\
	\qquad\qquad Sample $Y_i^{(j)}(z, t) = 1$ with probability $p_{iz}$\\
 	\qquad Causal effect estimate $\hat{\Delta_j}(t) = h(\textbf{Y}_1^{(j)}(t), \textbf{Y}_0^{(j)}(t))$ \\
\enspace Return $\Delta$
\end{tabbing}
\end{algorithm}

\section{Application on Kidney exchanges}
\subsection{Preliminaries}
\label{section:kidneys}
Kidney exchanges \citep{roth04} enable kidney transplantations when donors are incompatible with recipients.
In particular, a pair of a donor and a recipient who are incompatible can exchange a kidney transplant
with a pair of donor/recipient, provided that the donor from one pair can donate to the patient of the other.
Incompatilibility is determined by two medical tests. The first is one is a blood-type test between the donor and the recipient.
The second test is a sensitivity test which shows whether the recipient will accept or reject the kidney transplant 
from the donor. The statistics of these compatibilities are well studied and we will assume them to be known 
\footnote{For example, it is known that the probability that a random patient will reject a kidney of a random donor is about 0.11
and this is 3x as high when the recipient is a woman who has been pregnant and the donor is her spouse.}.
Typically, these exchanges involve 2 pairs due to logistical issues, however it is also common to perform 
\emph{cycle exchanges} in which a donor donates to the recipient of another pair, in sequence, until a loop is formed.
Multiple regional exchange programs currently operate in the US and the world, however, their expansion has 
hitherto been hindered by logistical and mechanism inefficiency issues. In specific, it has been reported 
that manipulation of centralized kidney exchange markets is possible and is performed by participating hospitals \citep{ashlagi2010mix}. Work in mechanism design has focused on mechanisms that resolve such incentives issues
\citep{ashlagi11, toulis11, ashlagi2010mix}.

The kidney exchange problem fits the framework of this paper as follows. First, we assume 
$N$ hospitals and we assume the existence of two mechanisms $\M{0}$ and $\M{1}$. 
The former, $\M{0}$, can be considered as the mechanism currently in practice whereas $\M{1}$ is a new
proposed mechanism under test. Agents are hospitals that are randomly assigned to participate in an exchange mechanism.
This exchange is multi-round (e.g. once per month) as it usually happens in practice.
At each round $t$, each hospital \emph{samples} a donor/patient pool of fixed size. This pool can be 
represented by a set of donor/patient pairs such that in each pair 
the donor is willing to donate to the patient but they are incompatible. Given such set, 
compatibilities can be determined by medical tests that are assumed \emph{common knowledge} i.e.,
hospitals cannot hide compatibilities between pairs, as that would be unethical and easy to uncover.
Thus, the sampled type of the hospital $\typei$ is simply the set of donor/recipient pairs that were 
sampled at round $t$. At each round, the hospital decides between two strategies: in the truthful strategy,
$\Yizt=1$, the hospital reports $\Rizt= \typei$. 
In the deviating strategy, $\Yizt=0$ and the hospital performs all possible matches among
its own pairs internally, and then reports the remainder $\Rizt = d(\typei)$. Given a pool of donor/patients $\typei$,
the function $d(.)$ is deterministic. Furthermore, as mentioned before, compatibility statistics are well-documented in the 
medical literature, and so the distributions of $\typei$ and $d(\typei)$ ($F$ and $G$ respectively) are assumed known\footnote{
For example, a pair with a donor with blood-type O is one that can possibly perform many exchanges, since 
O-donors can donate to all blood-types. Therefore, we expect more O-donors under distribution $F$ than distribution $G$, 
since when deviating, hospitals are more likely to match these ``good" pairs internally.}.

\subsection{Simulation Setup}
We perform simulation of two realistic stylized kidney exchange models that have been studied in literature.
Mechanism $\M{0}$ is the baseline mechanism and given a joint pool of hospital reports, computes a random maximum matching over all pairs. Mechanism $\M{1}$ applies the \emph{revelation principle} along with some more detailed allocations
in well-defined subgroups of donor/patient pairs\footnote{Briefly, pairs can be categorized as ``under-demanded", "over-demanded", ``reciprocal" and ``self-demanded". The compatibility networks within these groups vary significantly. 
The ``under-demanded" cannot be matched to each other and so the subgroup network is isolated.
The ``reciprocal" subgroup consists of two smaller groups that can be matched to each other and so the network is bipartite.
The ``self-demanded" is composed of four smaller groups that internally look like complete graphs.
These nuisances can be leveraged for the design of better allocation mechanisms than myopic maximum matchings.}.
Random graph theory has been leveraged to show that $\M{0}$ is vulnerable to deviating hospitals 
while in $\M{1}$ hospitals are better-off by being truthful \citep{toulis11}.

As a ground-truth model of agent strategic behavior, we adopt a multi-armed bandit formulation.
Specifically, we assume that hospitals try to maximize their utility (number of total pairs matched)
by using the \emph{uniform confidence bound} algorithm.
This algorithm has been widely used in practice as the simplest and most effective model of 
dynamic strategic behavior with bounded regret \citep{auer02}. The algorithm used to generate the dataset
is shown in Algorithm \ref{algo3}.

\begin{algorithm}[!h]
\caption{Simulation model of dynamic hospital behavior in kidney exchanges. Variable 
$n_{is}$ keeps track of how many times hospital $i$ has chosen strategy $s$ (1=truthful).
Variable $u_{is}$ keeps track of the average achieved utility of agent $i$ by playing strategy $s$.} 
\label{algo3}
\vspace*{5pt}
\begin{tabbing}
  \enspace For $\forall$ hospital $ i, \forall s \in \{0,1\}$ \\
  \qquad $u_{is}=0$ \\
  \qquad $n_{is}=1$ \\
   \enspace For $t = 1,2, \cdots T$ \\
  \qquad For $i = 1, 2, \cdots N$ \\
\qquad \qquad  $\typei \sim F$, sample the hospital's internal pairs \\
 \qquad \qquad  $s^* = \arg \max_{s \in \binary} \left ( u_{is} + \sqrt{\frac{2 \log t}{n_{is}}} \right )$ \\
 \qquad \qquad  $Y_i(z, t) = s^*$ \\
 \qquad \qquad  $n_{is^*} = n_{is^*} + 1$ \\
  \qquad\qquad    $R_i(Z_i, t)=\begin{cases}
    					\typei, & \text{if $Y_i(z, t)=1$} \nonumber \\
    					d(\typei), &  \text{ otherwise } 
  \end{cases}$ \\
 	\qquad Causal effect estimand $\Delta(t) = h(\textbf{Y}_1(t), \textbf{Y}_0(t))$ \\
\enspace Return $\Delta(t), \Rtobs, \Rcobs, \forall t$
\end{tabbing}
\end{algorithm}

\begin{figure}[h!]
\begin{center}
\includegraphics[scale=0.26]{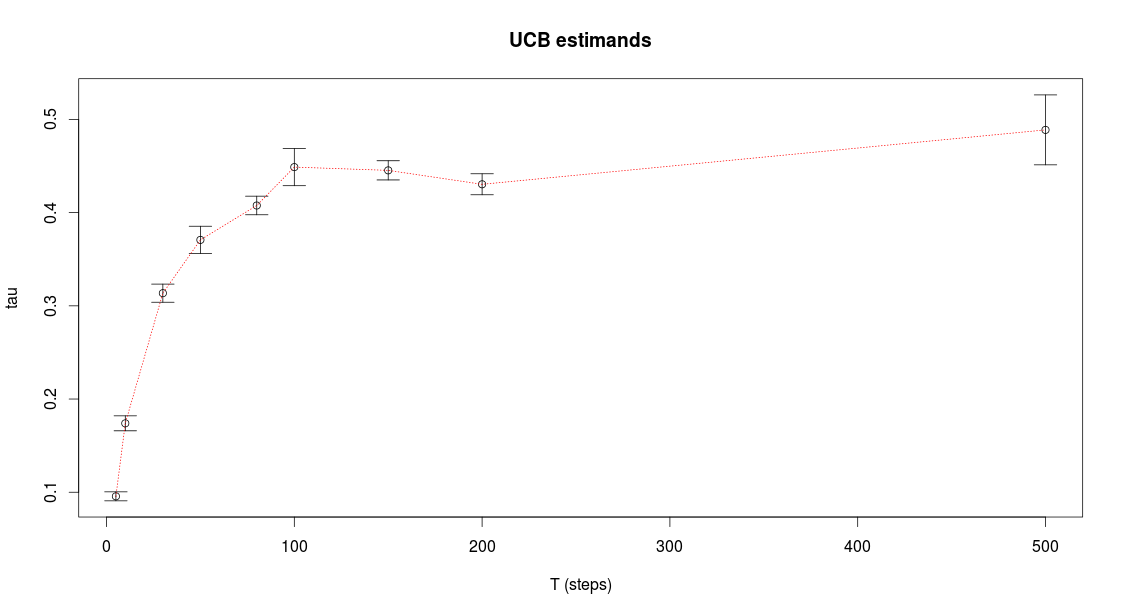}
\caption{Causal estimand at different rounds $t$ average over 100 independent runs.
The estimand is time-dependent indicating that the system reaches equilibrium over time.
}
\label{fig:ucb}
\end{center}
\end{figure}

The output of the simulation of Algorithm \ref{algo3}
are the causal estimand values at different rounds $t$ and 
the observed agent reports (($N/2 \times 1$) vectors of observed hospital reports for each $t$).
Our inference goal for both the empirical and the game-theoretic methods will be to estimate
the estimand values $\Delta(t)$ produced by the simulation, given the observed reports,
for \emph{two} specific rounds: $t_1=5$ will be round when the data are collected (observe 
agents' reports) and $t_2=T=100$ will be considered as the round where the system has reached 
equilibrium.
Figure \ref{fig:ucb} shows 100 independent runs of the simulation with the multi-armed bandit 
dynamic and the confidence bands of the respective estimand values $\Delta(t)$ for every round $t$.
Note that the causal effects to be estimated are time-dependent.
For example, at round $t=5$, the difference in incentives is around 0.1, specifically 0.5 for $\xcm$ and $0.4$ for $\rcm$,
which means that hospitals in $\xcm$ are 25\% more likely to play the truthful strategy ($t$)
compared to $\rcm$ at round $5$ and under the multi-armed bandit dynamic. 
For larger $t$ this value steadily increases until $0.44$ and then stabilizes (based on
additional experiments, we believe the seemingly linear trend at later time points, is an artifact of 
our simulation). The complication for causal inference is that unbiased estimates of 
incentives, taken at different timepoints, would be wildly different. 
This illustrates that estimation of mechanism effects needs to make the distinction 
between \emph{short-term} and \emph{long-term}. One key goal of this paper is to 
propose a methodology to estimate long-term effects by early experimental data.
In our simulation, we assume that data are collected at $t=5$ and we are interested in estimating 
the short term effect $\Delta(5)$ and long-term effect $\Delta(100)$.

\subsection{Estimation}
\label{section:estimation}
We compare between the empirical method that uses uniform priors and our method
which is using a game-theoretic prior. The former is straightforward and can be implemented
by following Algorithm \ref{algo1}.  The implementation of our method is a bit more involved 
as it first requires to compute the payoff functions $\Delta u(\cdot)$ as 
put forth in Equation \eqref{eq:utility} and under the exchangeability assumption (Assumption \ref{assumption:xchangeability}). In our simulation, this means that we have to calculate the payoff matrices
for 9 cases. These can be obtained through simulations.
The results over 10,000 mechanism simulations are shown in Table \ref{table:payoff}. For example, in $\rcm$ a truthful hospital will have an expected utility of $9.66$ matches when there are $5$ truthful hospitals out of $N=8$. Thus if this hospital were to deviate, 
it would obtain an expected utility of $10.69$ since the number of truthful hospital would decrease to 4.
Table \ref{table:payoff} shows that deviation is a dominant stratefy in $\rcm$ and 
truthful strategy is dominant strategy in $\xcm$.
\begin{table}[t]
\centering
\caption{Payoff matrices of $\rcm$ and $\xcm$. Each cell in the matrix shows the expected utility 
(average number of matched pairs) given (i) a mechanism, (ii) number of truthful hospitals in the mechanism and (iii) an agent strategy. The table shows that in $\rcm$ it is better for hospitals to deviate and in $\xcm$ it is better to 
be truthful.}
 \label{table:payoff}
\begin{tabular}{ccc | ccc}
\multicolumn{3}{c}{$\rcm$} & \multicolumn{3}{c}{$\xcm$} \\
\multicolumn{2}{c}{expected utility} &  & \multicolumn{2}{c}{expected utility} & \\
 truthful & deviating & \#truthful & truthful & deviating & \#truthful\\ 
- & 9.76 &   0 & - & 9.58 &   0 \\ 
   8.24 & 10.06 &   1 & 9.94 & 9.61 &   1 \\ 
   8.71 & 10.37 &   2 & 9.82 & 9.54 &   2 \\ 
   9.07 & 10.49 &   3 & 9.91 & 9.66 &   3 \\ 
   9.31 & 10.69 &   4 & 9.75 & 9.68 &   4 \\ 
   9.66 & 10.76 &   5 & 9.86 & 9.78 &   5 \\ 
   9.87 & 10.91 &   6 & 9.83 & 9.88 &   6 \\ 
   10.15 & 11.21 &   7 & 9.89 & 9.85 &   7 \\ 
   10.30 & - &   8 & 9.88 & - &   8 \\ 
\end{tabular}
\end{table}

Having obtained the payoff matrices, estimation through our model proceeds through the simple
Gibbs sampling procedure described in Algorithm \ref{algo3}.

%% RESULTS
\subsection{Results}
We conduct two experiments and for each experiment we collect agent reports 
at $t=5$ and wish to estimate short-term effects at $t=5$ and long-term effects for $T=100$.
The ground truth estimands have simulated values $\Delta(5)=0.1$ and $\Delta(100)=0.44$ 
over 500 independent simulation runs.

\begin{figure}[h!]
        \centering
        \begin{subfigure}[b]{0.4\textwidth}
               \includegraphics[scale=0.2]{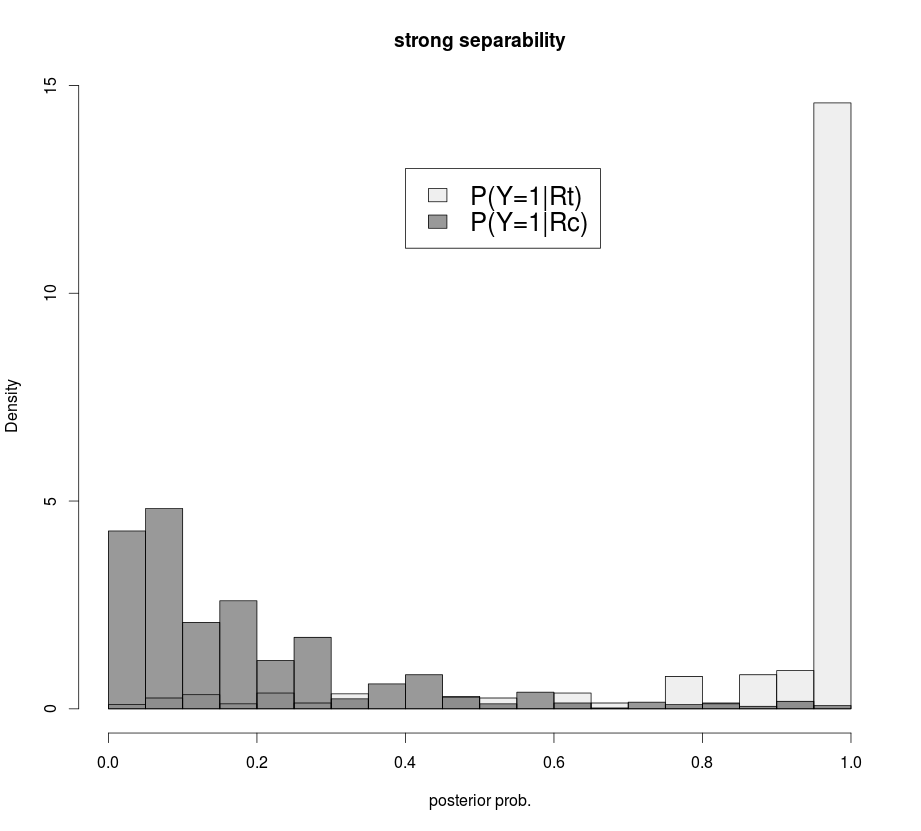}
                \caption{Conditional distribution of agent strategy conditioned on truthful report (light gray)
	or untruthful report (dark gray). }
                \label{fig:good-separation}
        \end{subfigure}%
        \quad\quad %add desired spacing between images, e. g. ~, \quad, \qquad etc.
        \quad \begin{subfigure}[b]{0.4\textwidth}
\includegraphics[scale=0.20]{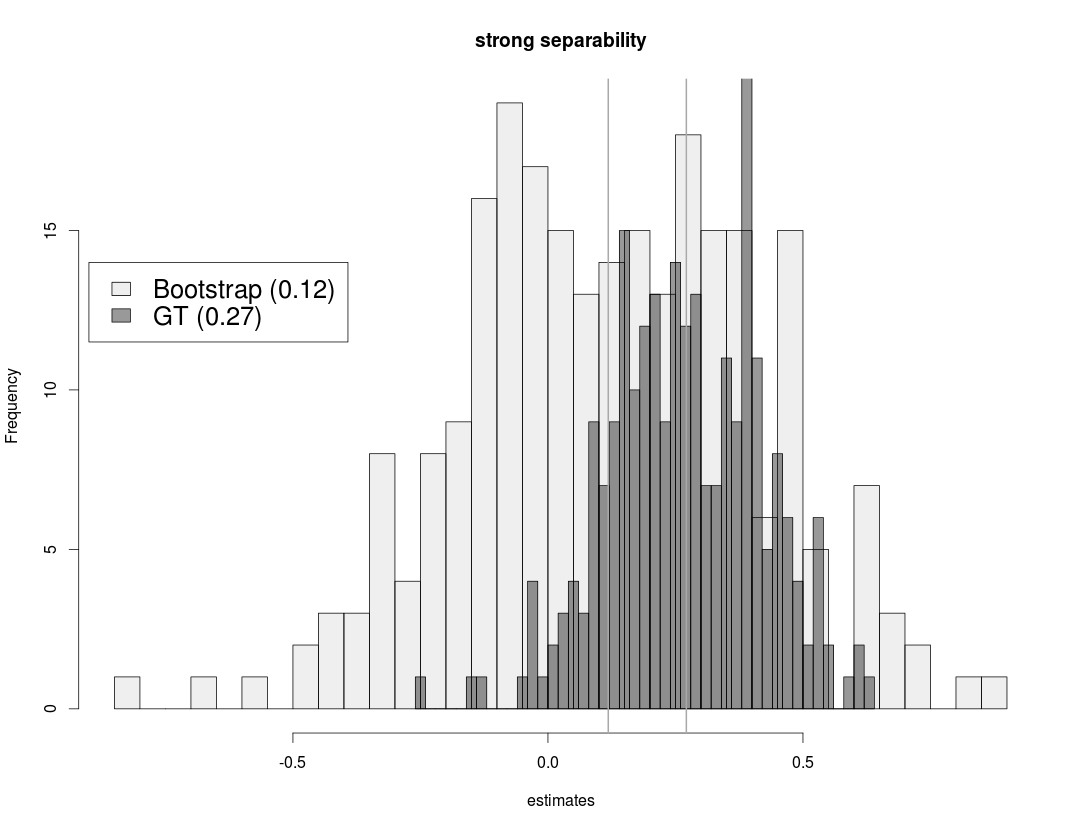}
                \caption{Causal effects estimates from the empirical method (light gray) and the game-theoretic (GT) method (dark gray).}
                \label{fig:estimates1}
        \end{subfigure}
        ~ %add desired spacing between images, e. g. ~, \quad, \qquad etc.
        \caption{Experiment in which agent reports are informative of agent strategies (strong separability).
	Ground truth estimands have simulated values $\Delta(5)=0.1$ and $\Delta(100)=0.44$.}
\end{figure}

In our first experiment, we work on a case in which agent reports are highly informative of the 
underlying agent behaviors. We refer to this case as ``strong separability" since the posterior distributions
$\Yizt |  \Rizt$ takes higher values (around 1) when the report is a truthful one and 
take small values (around 0) when the report is an untruthful one. In this cases the distributions look ``separated" as in Figure \ref{fig:good-separation}.

Figure \ref{fig:estimates1} shows histograms of the estimates from the empirical method and 
the game-theoretic method. We can see that the former estimates $\Delta(5)$ at 0.12. Visual inspection 
of the estimates shows that the method performs well under the informative agent reports (strong separability).
This is expected, since the likelihood gives plenty of information about the missing agent strategies. However, while 
the estimate (under the exchangeability assumption) is unbiased for the true effect $\Delta(5)$ at round $t=5$,
it is still biased for the long-term effect $\Delta(100)$. The game-theoretic method makes a compromise between the two 
estimands. The estimates are centered around  0.27 and they clearly biased for the effect at $t=5$.
However, they are also shrinked towards the long-term effect $\Delta(100)$ which indicates that the payoff matrix is able to capture the dynamic evolution of the system to a certain extent.

\begin{figure}[h!]
        \centering
        \begin{subfigure}[b]{0.4\textwidth}
            \includegraphics[scale=0.18]{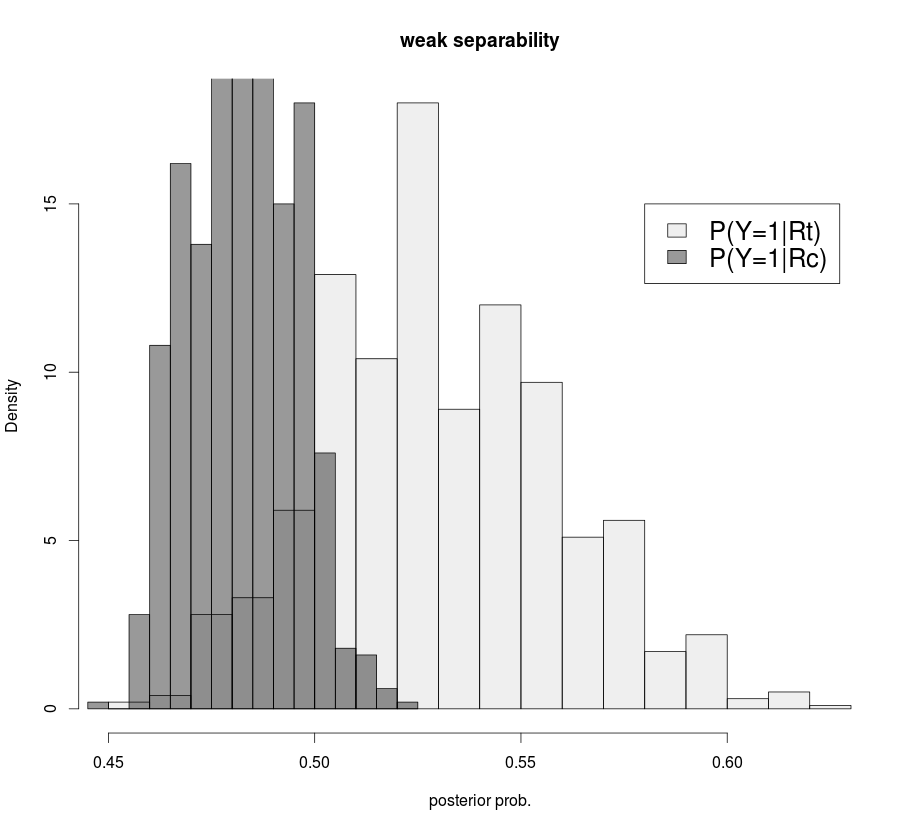}
                \caption{Conditional distribution of agent strategy conditioned on truthful report (light gray)
	or untruthful report (dark gray). }
                \label{fig:bad-separation}
        \end{subfigure}%
        \quad\quad %add desired spacing between images, e. g. ~, \quad, \qquad etc.
        \quad \begin{subfigure}[b]{0.4\textwidth}
\includegraphics[scale=0.22]{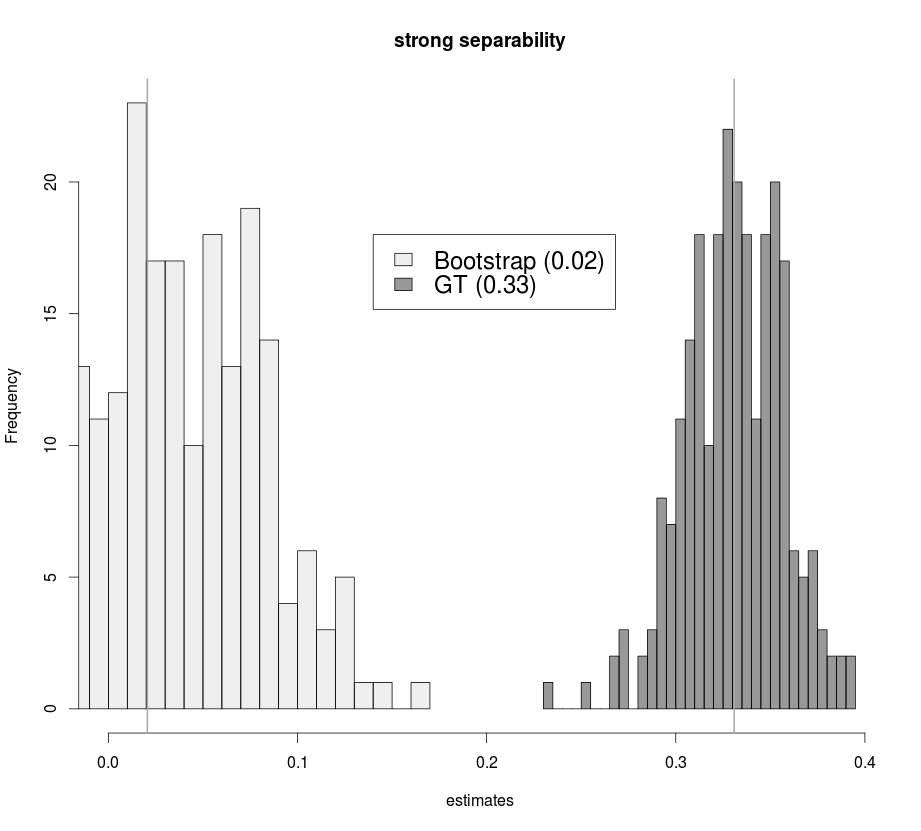}
                \caption{Causal effects estimates from the empirical method (light gray) and the game-theoretic (GT) method (dark gray).}
                \label{fig:estimates2}
        \end{subfigure}
        ~ %add desired spacing between images, e. g. ~, \quad, \qquad etc.
        \caption{Experiment in which agent reports are informative of agent strategies (strong separability).
	Ground truth estimands have simulated values $\Delta(5)=0.1$ and $\Delta(100)=0.44$.}
\end{figure}

In our second experiment, we work on a case in which agent reports are not informative about the 
underlying agent behaviors. We refer to this case as ``weak separability" since the posterior distributions
$\Yizt |  \Rizt$ takes values around 0.5 for both truthful and untruthful reports so that 
there is practically no information about strategies given observed agents' reports. The respective
distributions are shown in Figure \ref{fig:bad-separation}.

Figure \ref{fig:estimates2} shows histograms of the estimates from the empirical method and 
the game-theoretic method for the weak separability case.
 We can see that the former estimates $\Delta(5)$ at 0.02. 
In that case, we observe a breakdown of the empirical method since the estimates are centered around zero.
This is because the reports are not informative (almost random guesses) about the strategies and so the empirical method is using 
only the information on the prior to make inference about incentives. However, since the empirical method 
is assuming a uniform prior, the overall procedure will deduce no difference in incentives.
In constrast, the game-theoretic method is actually giving higher estimates 
for the overall difference in incentives since it is based only on the equilibrium model of the prior.
Thus, the overall estimates are even higher than before (average around 0.33) as the equilibrium 
model shrinks the estimates towards long-term effects.

%%%   END OF ADDED STUFF
\section{Discussion}
The evaluation of mechanisms is critical in numerous socioeconomic problems.
However, this is technically challenging because multi-agent systems are dynamic by nature
and estimation should be performed with respect to an equilbrium state of the system.
Furthermore, in estimating effects of incentives, there are additional challenges as agent strategies,
which are the main potential outcomes of interest, are typically never observed. 
For the former, we use agent reports and further distributional assumptions to obtain 
likelihoods of strategies given observed reports. For the latter, we propose  a prior on agent strategies that is based on a quantal response equilibrium model. 
In a simulated study, this was shown to shrink towards long-term effects, thus 
offering improved inference over methods that don't consider such priors.

There are multiple ways that this work could be further improved.
First, the empirical method we described is by now means the only way that could be used to perform causal
inference. A fully Bayesian model would also be a good choice. However, this work hints that
any model that is ignoring the game-theoretic aspect of the potential outcomes (agent behaviors)
will be inadequate to capture the time dependence of the causal estimand. Second, the choice of the $\beta$ parameter in the game-theoretic prior is crucial and how it was set, was not sufficiently justified. In practice, 
$\beta$ was set based on a 
heuristic calculation and experimental results. Future work would be benefited by a more principled 
way to set such hyperparameters. Third, we offered limited discussion of our assumptions 
(best-response and exchangeability). Several violations of these assumptions yield more realistic 
and particularly interesting situations. For example, in case of substitution effects i.e., cases where 
agents can switch between mechanisms (e.g. assume that a mechanism is a mode of transportation), 
it is no longer possible to model the two potential outcome vectors independently. Last but not least, 
agent interactions (e.g. information sharing, communication, collusion) will require more sophisticated models.

\bibliographystyle{natbib}
\bibliography{paper-ref}

\begin{thebibliography}{17}
\expandafter\ifx\csname natexlab\endcsname\relax\def\natexlab#1{#1}\fi

\bibitem[{Ashlagi et~al.(2010)Ashlagi, Fischer, Kash \&
  Procaccia}]{ashlagi2010mix}
\textsc{Ashlagi, I.}, \textsc{Fischer, F.}, \textsc{Kash, I.} \&
  \textsc{Procaccia, A.~D.} (2010).
\newblock Mix and match.
\newblock In \textit{Proceedings of the 11th ACM conference on Electronic
  commerce}. ACM.

\bibitem[{Ashlagi \& Roth(2011)}]{ashlagi11}
\textsc{Ashlagi, I.} \& \textsc{Roth, A.} (2011).
\newblock Individual rationality and participation in large scale,
  multi-hospital kidney exchange.
\newblock In \textit{Proceedings of the 12th ACM conference on Electronic
  commerce}. ACM.

\bibitem[{Athey \& Nekipelov(2010)}]{athey2010}
\textsc{Athey, S.} \& \textsc{Nekipelov, D.} (2010).
\newblock A structural model of sponsored search advertising auctions.
\newblock In \textit{Sixth Ad Auctions Workshop}.

\bibitem[{Auer et~al.(2002)Auer, Cesa-Bianchi \& Fischer}]{auer02}
\textsc{Auer, P.}, \textsc{Cesa-Bianchi, N.} \& \textsc{Fischer, P.} (2002).
\newblock Finite-time analysis of the multiarmed bandit problem.
\newblock \textit{Machine learning} \textbf{47}, 235--256.

\bibitem[{Bottou et~al.(2012)Bottou, Peters, Qui{\~n}onero-Candela, Charles,
  Chickering, Portugualy, Ray, Simard \& Snelson}]{bottou2012}
\textsc{Bottou, L.}, \textsc{Peters, J.}, \textsc{Qui{\~n}onero-Candela, J.},
  \textsc{Charles, D.~X.}, \textsc{Chickering, D.~M.}, \textsc{Portugualy, E.},
  \textsc{Ray, D.}, \textsc{Simard, P.} \& \textsc{Snelson, E.} (2012).
\newblock Couterfactual reasoning and learning systems.
\newblock \textit{arXiv preprint arXiv:1209.2355} .

\bibitem[{Clarke(1971)}]{clarke71}
\textsc{Clarke, E.~H.} (1971).
\newblock Multipart pricing of public goods.
\newblock \textit{Public choice} \textbf{11}, 17--33.

\bibitem[{Goeree et~al.(2003)Goeree, Holt \& Palfrey}]{goeree03}
\textsc{Goeree, J.~K.}, \textsc{Holt, C.~A.} \& \textsc{Palfrey, T.~R.} (2003).
\newblock Risk averse behavior in generalized matching pennies games.
\newblock \textit{Games and Economic Behavior} \textbf{45}, 97--113.

\bibitem[{Groves(1973)}]{groves73}
\textsc{Groves, T.} (1973).
\newblock Incentives in teams.
\newblock \textit{Econometrica: Journal of the Econometric Society} , 617--631.

\bibitem[{Heckman \& Vytlacil(2005)}]{heckman05}
\textsc{Heckman, J.~J.} \& \textsc{Vytlacil, E.} (2005).
\newblock Structural equations, treatment effects, and econometric policy
  evaluation1.
\newblock \textit{Econometrica} \textbf{73}, 669--738.

\bibitem[{Lubin \& Parkes(2009)}]{lubin09}
\textsc{Lubin, B.} \& \textsc{Parkes, D.~C.} (2009).
\newblock Quantifying the strategyproofness of mechanisms via metrics on payoff
  distributions.
\newblock In \textit{Proceedings of the Twenty-Fifth Conference on Uncertainty
  in Artificial Intelligence}. AUAI Press.

\bibitem[{McKelvey \& Palfrey(1995)}]{mckelvey95}
\textsc{McKelvey, R.~D.} \& \textsc{Palfrey, T.~R.} (1995).
\newblock Quantal response equilibria for normal form games.
\newblock \textit{Games and economic behavior} \textbf{10}, 6--38.

\bibitem[{Ostrovsky \& Schwarz(2011)}]{ostrovsky2011}
\textsc{Ostrovsky, M.} \& \textsc{Schwarz, M.} (2011).
\newblock Reserve prices in internet advertising auctions: A field experiment.
\newblock In \textit{Proceedings of the 12th ACM conference on Electronic
  commerce}. ACM.

\bibitem[{Roth et~al.(2004)Roth, S{\"o}nmez \& {\"U}nver}]{roth04}
\textsc{Roth, A.~E.}, \textsc{S{\"o}nmez, T.} \& \textsc{{\"U}nver, M.~U.}
  (2004).
\newblock Kidney exchange.
\newblock \textit{The Quarterly Journal of Economics} \textbf{119}, 457--488.

\bibitem[{Rubin(1974)}]{rubin74}
\textsc{Rubin, D.~B.} (1974).
\newblock Estimating causal effects of treatments in randomized and
  nonrandomized studies.
\newblock \textit{Journal of educational Psychology} \textbf{66}, 688.

\bibitem[{Rubin(1978)}]{rubin78}
\textsc{Rubin, D.~B.} (1978).
\newblock Bayesian inference for causal effects: The role of randomization.
\newblock \textit{The Annals of Statistics} , 34--58.

\bibitem[{Toulis \& Parkes(2011)}]{toulis11}
\textsc{Toulis, P.} \& \textsc{Parkes, D.~C.} (2011).
\newblock A random graph model of kidney exchanges: efficiency,
  individual-rationality and incentives.
\newblock In \textit{Proceedings of the 12th ACM conference on Electronic
  commerce}. ACM.

\bibitem[{Vickrey(1961)}]{vickrey61}
\textsc{Vickrey, W.} (1961).
\newblock Counterspeculation, auctions, and competitive sealed tenders.
\newblock \textit{The Journal of finance} \textbf{16}, 8--37.

\end{thebibliography}

\end{document}